\documentclass[12pt]{article}

\usepackage[T1]{fontenc}
\usepackage[utf8]{inputenc}
\usepackage{float}
\usepackage{amsmath}
\usepackage{graphicx}
\usepackage{amsmath}
\usepackage{amssymb}
\usepackage{framed}
\usepackage{subfigure}
\usepackage[italian, english]{babel}
\usepackage{enumitem}
\usepackage{longtable}
\usepackage{mdwlist}
\usepackage[a4paper]{geometry}
\usepackage{tabularx}
\usepackage{booktabs}
\usepackage{lscape}

\usepackage{parskip}

\newcommand{\qed}{\hspace*{\fill} $\bigstar$\medskip}

\newtheorem{theorem}{Theorem}[section]

\DeclareMathOperator*{\argmin}{arg\,min}

\def \N {\mathbb N}

\def \s {\sigma}

\def \a {\alpha}
\def \g {\gamma}

\def \cav {\mathcal C}
\def \met {\mathcal M}
\def \gre {\mathcal G}

\usepackage{tikz}

\author{
Benedetto Scoppola\\
Dipartimento di Matematica\\
Universit\`a degli Studi di Roma ``Tor Vergata''
\and
Alessio Troiani\\
Dipartimento di Matematica e Informatica ``Ulisse Dini''\\
Universit\`a degli Studi di Firenze
}
\title{Gaussian Mean Field Lattice Gas}
\date{\today}

\begin{document}

\maketitle

\begin{abstract}
	We study rigorously
	a lattice gas version of the Sherrington-Kirckpatrick spin glass model.
	In discrete optimization literature this problem is known as Unconstrained
	Binary Quadratic Programming (UBQP) and it belongs to the class NP-hard.
	We prove that the fluctuations of the ground state energy tend to vanish
	in the thermodynamic limit, and we give a lower bound of such ground state
	energy.
	Then we present an heuristic algorithm, based on a probabilistic cellular
	automaton, which seems to be able to find configurations with energy very close to
	the minimum, even for quite large instances.
\end{abstract}

\section{Introduction}
\label{Intro}

In this paper we give some rigorous results about the ground state of the system described
by the Hamiltonian
\begin{align}\label{eq:hamiltonian}
H(\eta)=\frac{1}{\sqrt N}\sum_{i,j=1}^N J_{ij}\eta_i\eta_j
\end{align}
where $\eta\in\{0,1\}^N$, the coefficients $J_{ij}$ are i.i.d. Gaussian random variables with
mean 0 and variance 1, and the sum runs over all pairs $i,j$, giving a
non vanishing contribution also for the terms $J_{ii}$.
Even if the mean field interaction rules out any geometry in the system, we
use the name ``lattice gas'' since the
variables in the system are defined in $\{0,1\}$ instead of in $\{-1,1\}$.

Although the Hamiltonian \eqref{eq:hamiltonian}
defined in terms of variables $\s\in\{-1,1\}^N$ is one of system with the wider bibliography in
theoretical physics of the last 40 years, under the name of Sherrington-Kirkpatrick ($SK$) model,
its lattice-gas counterpart did not
receive the same attention in the mathematical physics literature:
to the best of our knowledge only
Russo in \cite{Russo} studied a similar problem, adding to the Hamiltonian
a term related to an external field $h$.

From the point of view of the combinatorial optimization, on the other side,
the problems of finding the ground state of the system described in (\ref{eq:hamiltonian})
in the cases $\eta\in\{0,1\}^N$ and $\s\in\{-1,1\}^N$ are
both widely studied. They are both NP-hard problems, and hence they are polynomially equivalent,
but they are known under different names: the search
of the ground state of the $\{-1,1\}$ SK system is equivalent, from an algorithmic
point of view, to a max cut problem (see for instance
Rinaldi et al in \cite{DJMRR}). This can be stated in the following way:
given a weighted complete graph with $N$ vertices and edge weights $w_{ij}=-2J_{ij}$ (the self interaction
terms $w_{ii}=-J_{ii}$ are in this case unessential), find
the partition of the vertex set in two disjoint sets $I_1$ and $I_2$ such that
$I_1\cup I_2=\{1,2,...,N\}$ and the sum $-\sum_{i_1\in I_1, i_2\in I_2} w_{i_1,i_2}$
is maximum.
The problem studied in this paper, in which $\eta\in\{0,1\}^N$,
is known in combinatorial optimization under the name \emph{Unconstrained Binary Quadratic
Programming} (UBQP) problem, and it corresponds to the search, in the same graph mentioned before,
of the subset $I\subset \{1,2,...,N\}$ such that $\sum_{i,j\in I} w_{ij}$ is maximum.
The literature about this problem from a combinatorial point of view is huge, see
for instance \cite{KHGLL} and reference therein.
In recent times Lasserre \cite{Lass} has outlined that the two problems may be mapped
one onto the other, and therefore many results (mainly upper and lower bounds on the
minimum) on the max cut problem may be used in the context of the UBQP.
The connection between max cut and UBQP is obviously given by the
change of variables defined by
$$
\eta=\frac{\sigma +1}{2}\qquad\sigma = 2\eta -1
$$
Hence the $SK$ hamiltonian
$$
H(\sigma)=\frac{1}{\sqrt N}\sum_{i,j=1}^N J_{ij}\sigma_i\sigma_j
$$
is equivalent to 
\begin{align}\label{eq:hamiltonian2}
H(\eta)=\frac{4}{\sqrt N}\sum_{i,j=1}^N J_{ij}\eta_i\eta_j-\frac{2}{\sqrt N}\sum_{i,j=1}^N J_{ij}(\eta_i+\eta_j)
\end{align}
Since we are interested, in this paper, in the properties
of the ground state as a random function of the realization of the $J_{ij}$, the
extra linear term, that is dependent on the whole set of $J_{ij}$,
and hence gives a contribution to the hamiltonian of the same order of the
quadratic term, makes the two problems, 
max cut and UBQP, not equivalent. In other words, for fixed (quenched) values of $J_{ij}$
the configuration that minimizes \eqref{eq:hamiltonian} has nothing to share with the minimizer of
\eqref{eq:hamiltonian2}. This is apparent also from the fact that the average magnetization of the
minimizer $\sigma^{min}$ of the SK model vanishes in the thermodynamic limit, giving
fot the corresponding $\eta^{min}$ an average density $\frac{1}{N}\sum_{i=1}^N\eta_i^{min}=
\frac{1}{2}$,
while the average density of the minimizer of the UBQP problem is larger (see below).
 
However it is immediate to realize numerically, as we will show in 
Section~\ref{sec:a_pca_based_algorithm_to_find_the_minima_of_h}, that 
the system should share with the SK model many features. One of the most
important feature that we expect for this system is the existence, almost surely, of a well
defined minimum energy per particle in the thermodynamic limit. In SK model the proof of the existence of such minimum energy per particle is based on existence of the limit of the free energy which, in its turn, is proven using a subadditivity argument \cite{GT}. 
The proof of the subadditivity of the free energy uses the property $\sigma^{2} = 1$. However,
 in our case the dynamical variables $\eta$ are idempotent: $\eta^2=\eta$. Hence
the proof of the subadditivity of the free energy is not obvious. This will be the subject
of a further effort, but at the moment we state as a conjecture the following claim.

\vglue.3truecm
{\bf Conjecture}

{\it For the UBQP problem in the
gaussian case, define the quantities $M_N$ and $m_N$ and the configuration $\eta^*$ by

$$\min_{\eta\in\{0,1\}^N}H(\eta):= H(\eta^*):= -M_N:=-m_NN
$$

Clearly $M_N$, $m_N$ and the configuration $\eta^*$
realizing the minimum depend on the disorder, i.e.,
on the entries of the matrix $J_{ij}$. 

There exist $\bar m>0$ and $0<\bar\alpha<1$ such that the following limits hold almost surely

$$
\lim_{N\rightarrow\infty}\frac{M_N}{N}=\lim_{N\rightarrow\infty}m_N=\bar m\qquad
\lim_{N\rightarrow\infty}\frac{\sum_{i=1}^N\eta^*_{i}}{N}=\bar\alpha
$$
}

From a numerical point of view (see below) reasonable estimates of $\bar m$  and $\bar\alpha$ are
$\bar m\approx .42$ and $\bar\alpha\approx .62$.

In this paper we prove two preliminary results in the direction of the proof of the
conjecture above. In Section~\ref{flu} we prove that the fluctuations of $m_N$
around its average value ${E}(m_N)$ tend to zero in the thermodynamic limit, where 
$E$ denotes the expected value with respect to the disorder $J$.

The techniques we use to achieve this result are mutuated from \cite{BGP}. More precisely
the result we prove can be stated as follows.
Take a sequence
of realizations of the disorder $J_{ij}^{(N)}$ indexed by the volume $N$.
Then $m_{N}$
tends to $E(m_N)$ almost surely as $N\rightarrow\infty$.

In Section~\ref{low} we give two estimates of the values $E(m_N)$ that are uniform in $N$.
The first estimate is a simple annealed computation. We define for $m>0$
the random variable $\nu(m)$ as the number
of configurations with energy below $-mN$. 
Evaluating $E(\nu(m))$ we obtain $E(m_N)<.801$. 
For the second estimate we
compute $E(\nu(m))$ conditioning on the disorder $J$ to 
be typical, i.e., imposing that the deviation of $H({\bf 1}):=\sum_{i,j}J_{i,j}$
is of the order of $\sqrt{N}$. 

This conditioning gives a remarkable improvement of the estimate: we obtain in this case
$E(m_N)<.562$. This proves explicitly that for $.562<m<.801$ the main contribution to the expected value
of $\nu(m)$, that is exponentially large in $N$, is given by non-typical realizations of the disorder $J_{ij}$. 

In the second part of this paper, namely in Section~\ref{sec:a_pca_based_algorithm_to_find_the_minima_of_h}, we present a parallel algorithm, inspired by the recent results on
the Gibbs sampling of spin systems by means of Probabilistic Cellular  Automata (see \cite{ISS} \cite{DSS} \cite{PSS}) and we discuss its promising performances.

\section{Small fluctuation of the minimum energy per particle}
\label{flu}

To lighten the notation we drop, in this section, the subscript $N$ and we write 
$$M=\min_{\eta\in\{0,1\}^N}H(\eta)$$
We define $\Delta :=M-E(M)$.
\begin{theorem}\label{thm_small_fluctuations}
For a suitable constant
$C>0$ we have that for all $z>0$
\begin{equation}\label{2}
P(|\Delta|>Nz)\le e^{-CNz^2}
\end{equation}

\end{theorem}

\emph{Proof}:

We consider $P(\Delta>Nz)$. The evaluation of $P(\Delta<-Nz)$ is done in the same way.

By exponential Markov inequality we have that for all $t>0$
\begin{equation}\label{3}
P(\Delta>Nz)\le e^{-tNz}E(e^{t\Delta})
\end{equation}
Following the arguments introduced in \cite{BGP}, we fix an arbitrary positive integer
$K$ and we observe that each gaussian variable $J_{ij}$
is distributed as
the sum of $K$ i.i.d. gaussian variables $J_{ij}^{(l)},\ l=1,...,k$, of average zero and variance $1/K$.
We fix an order on the set of $KN^2$ gaussian variables
obtained by this procedure and we label with a single index $k$ each of them,
that are therefore indicated as $J_k$.
In this way to each $k$ corresponds a pair of indices $i(k),j(k)$.
With this notations we may rewrite
the minimum $M$ as
\begin{equation}\label{minimo}
M=\min_\eta\frac{1}{\sqrt N}\sum_{k=1}^{KN^2}\eta_{i(k)}\eta_{j(k)}J_k
\end{equation}
Denote now by $E_I(\cdot)$, with $I\subset\{1,2,...,KN^2\}$, the expectation
with respect to the $J_k$'s with $k\in I$.
It is immediate to realize that
$$
\Delta=M-E_{\{1\}}(M)+E_{\{1\}}(M)-E_{\{1,2\}}(M)+E_{\{1,2\}}(M)-
E_{\{1,2,3\}}(M)+...-E(M)
$$
Call $\Delta_i=E_{\{1,2,...,i-1\}}(M)-E_{\{1,2,...,i\}}(M)$. Clearly $\Delta=\sum_i\Delta_i$ and hence
$e^{t\Delta}=\prod_{i=1}^{KN^2}e^{t\Delta_i}$.
We have then
$$E(e^{t\Delta})=E\left(\prod_{i=1}^{KN^2}e^{t\Delta_i}\right)$$
We want now give an estimate of this expression iteratively, showing that
$$
E\left(\prod_{i=l}^{KN^2}e^{t\Delta_i}\right)\le E\left(\prod_{i=l+1}^{KN^2}e^{t\Delta_i}\right)L(K,N)
$$
with $L(K,N)\le e^{\frac{Ct^2}{NK}}$ for $K$ large enough.

\noindent
For this purpose, we write
$$
E\left(\prod_{i=l}^{KN^2}e^{t\Delta_i}\right)=E\left(\left(\prod_{i=l+1}^{KN^2}e^{t\Delta_i}\right)
e^{t\Delta_l}\right)=$$
$$
=E_{\{l+1,...,KN^2\}}\left(\left(\prod_{i=l+1}^{KN^2}e^{t\Delta_i}\right)
E_{\{l\}}(e^{t\Delta_l})\right)
$$
Hence we have to consider
\begin{equation}\label{stimaslice}
E_{\{l\}}(e^{t\Delta_l})=1+\frac{t^2}{2}E_{\{l\}}(\Delta^2_l)+R_3(\Delta_l)
\end{equation}
where
$$R_3(\Delta_l)=\frac{t^3}{3!}E_{\{l\}}\left(e^{\tilde t\Delta_l}\ \Delta_l^3\right)
\qquad 0\le{\tilde t}\le t
$$
and we have used the fact that
$E_{\{l\}}(\Delta_l)=0$.

In order to evaluate \eqref{stimaslice} it is useful to compare the minimum $M$
defined in \eqref{minimo} with the quantity
\begin{equation}\label{minimol}
{\widetilde M}_l=
\min_\eta{\widetilde H}_l(\eta)
=\min_\eta\frac{1}{\sqrt N}\sum_{k\ne l}\eta_{i(k)}\eta_{j(k)}J_k
\end{equation}
i.e., the minimum of the quadratic form obtained by neglecting the Gaussian
random variable $J_l$.
We prove the following result.

{\bf Lemma} (estimate of $|M-{\widetilde M}_l|$).

\begin{equation}\label{diff}
\sqrt N|M-{\widetilde M}_l|\le|J_l|
\end{equation}

\emph{Proof.}
We first fix some notations. Let us call $\bar\eta$ the configuration such that $H(\bar\eta)=M$, and
$\tilde\eta$ the configuration such that ${\widetilde H}_l(\tilde\eta)={\widetilde M}_l$.
Moreover, we will say that $n\in \eta$ if $\eta_{i(n)}\eta_{j(n)}=1$, and we will call

$$D=
\sum_{k\ne l}\bar\eta_{i(k)}\bar\eta_{j(k)}J_k-
\sum_{k\ne l}\tilde\eta_{i(k)}\tilde\eta_{j(k)}J_k
$$

The proof of the lemma can be done considering four possible cases

Case 1: $l\notin \bar\eta,\ l\notin\tilde\eta$

In this case
$$
{\sqrt N}M=\sum_{k\ne l}\bar\eta_{i(k)}\bar\eta_{j(k)}J_k,\qquad
{\sqrt N}{\widetilde M}_l=\sum_{k\ne l}\tilde\eta_{i(k)}\tilde\eta_{j(k)}J_k
$$

Hence it has to be $M={\widetilde M}_l$, which implies \eqref{diff}.

Case 2: $l\in \bar\eta,\ l\in\tilde\eta$

In this case
$$
\sqrt N M=\sum_{k\ne l}\bar\eta_{i(k)}\bar\eta_{j(k)}J_k +J_l,\qquad
\sqrt N{\widetilde M}_l=\sum_{k\ne l}\tilde\eta_{i(k)}\tilde\eta_{j(k)}J_k
$$

In order to realize the minimum $M$ in $\bar\eta$ and ${\widetilde M}_l$ in $\tilde\eta$ it has to be true that
$D=0$ and therefore $\sqrt N M=
\sqrt N{\widetilde M}_l+J_l$,
which implies \eqref{diff}.

Case 3: $l\in \bar\eta,\ l\notin\tilde\eta$

In this case
$$
\sqrt N M=\sum_{k\ne l}\bar\eta_{i(k)}\bar\eta_{j(k)}J_k +J_l,\qquad
\sqrt N {\widetilde M}_l=\sum_{k\ne l}\tilde\eta_{i(k)}\tilde\eta_{j(k)}J_k
$$
Since ${\bar\eta}\neq{\tilde\eta}$ it has to be true that
$D>0$, because it is convenient to switch from $\bar\eta$ to $\tilde\eta$
in order to minimize ${\widetilde M}_l$. Moreover it must be 
$J_l<0$ and $|J_l|>D$, because it is necessary that
$M<{\widetilde M}_l$.
Hence $\sqrt N M-\sqrt N {\widetilde M}_l=D+J_l$, which implies \eqref{diff}.

Case 4: $l\notin \bar\eta,\ l\in\tilde\eta$

In this case
$$
\sqrt N M=\sum_{k\ne l}\bar\eta_{i(k)}\bar\eta_{j(k)}J_k ,\qquad
\sqrt N {\widetilde M}_l=\sum_{k\ne l}\tilde\eta_{i(k)}\tilde\eta_{j(k)}J_k
$$
Again, in order to realize the minimum in $M$ and in ${\widetilde M}_l$ it has to be true that
$D>0$, because it is convenient to switch from $\bar\eta$ to $\tilde\eta$
in order to minimize ${\widetilde M}_l$. In this case, however,
$J_l>D$, because in the opposite case it would be convenient to put
$\bar\eta=\tilde\eta$ in order to minimize $M$.  Since in this case
$\sqrt N M- \sqrt N{\widetilde M}_l=D$, $J_l>D$ implies \eqref{diff}, and the Lemma is proven.

We can now use the lemma in order to evaluate $E_{\{l\}}(\Delta^2_l)$.
We have, indeed
$$
|\Delta_l|=\left|E_{\{1,2,...,l-1\}}(M)-E_{\{1,2,...,l\}}(M)\right|=
$$
\begin{equation}\label{diftilde}
=
\left|E_{\{1,2,...,l-1\}}(M-{\widetilde M}_l)-E_{\{1,2,...,l\}}(M-{\widetilde M}_l)
\right|
\end{equation}
The second line is due to the linearity of the expected value
and to the fact that $E_{\{1,2,...,l\}}({\widetilde M}_l)=
E_{\{1,2,...,l-1\}}({\widetilde M}_l)$, since ${\widetilde M}_l$ does not depend on $J_l$.
We have then
$$
E_{\{l\}}(\Delta^2_l)=
E_{\{l\}}\left[(E_{\{1,2,...,l-1\}}(M-{\widetilde M}_l)-
E_{\{1,2,...,l\}}(M-{\widetilde M}_l))^2\right]\le
$$
$$
\le 2E_{\{l\}}\left[E^2_{\{1,2,...,l-1\}}(|M-{\widetilde M}_l|) +
E^2_{\{1,2,...,l\}}(|M-{\widetilde M}_l|)\right]\le
$$
$$
\le \frac{2}{N} \left[ E_{\{l\}}(|J_l|^2)+E^2_{\{l\}}(|J_l|)\right]\le \frac{4}{NK}
$$
where in the last line we used the lemma above and the fact that the variance of
$J_l$ is $1\over K$.

For the term $R_3(\Delta_l)$, using again the lemma above together with \eqref{diftilde}
and $ 0\le{\tilde t}\le t$, the following inequalities hold,
$$R_3(\Delta_l)=\frac{ t^3}{3!}E_{\{l\}}\left(e^{\tilde t\Delta_l}\ \Delta_l^3\right)\le
\frac{ t^3}{3!}E_{\{l\}}\left(e^{t|\Delta_l|}\ |\Delta_l^3|\right)\le
\frac{C t^3}{3!N^{3/2}}E_{\{l\}}\left(e^{ct|J_l|/\sqrt N}\ |J_l^3|\right)
$$
with $C$ and $c$ suitable constants.
Computing the expectation, recalling that the variance of
$J_l$ is $1\over K$, yields
$$R_3(\Delta_l)\le
\frac{C t^3}{3!(NK)^{3/2}}e^{\frac{ct^2}{KN}}
$$

These two results allow us to bound \eqref{stimaslice}
\begin{equation}\label{stimaslice2}
E_{\{l\}}(e^{t\Delta_l})=1+\frac{t^2}{2}E_{\{l\}}(\Delta^2_l)+R_3(\Delta_l)\le 1+
\frac{C_1t^2}{2NK} + \frac{C_2 t^3}{3!(NK)^{3/2}}e^{\frac{ct^2}{NK}}\le e^{\frac{Ct^2}{NK}}
\end{equation}
for suitable constants $C_1,C_2$ and $C$, and for $K$ large enough.

Thanks to these results we get, iterating on all slices,
$$
E(e^{t\Delta})\le \left(e^{\frac{Ct^2}{NK}}\right)^{N^2K}\le e^{CNt^2}
$$
and hence we get that for all $t>0$
\begin{equation}\label{10}
P(\Delta>Nz)\le e^{-tNz}e^{CNt^2}.
\end{equation}

Choosing  $t=\frac{z}{2C}$ in \eqref{10} we get
\begin{equation}\label{11}
P(\Delta>Nz)\le e^{-N\frac{z^2}{4C}}
\end{equation}
which completes the proof of the theorem.

\section{Lower estimates of the ground state energy.}
\label{low}

Consider the Hamiltonian defined in \eqref{eq:hamiltonian}
and call $\nu(m)$ the random variable (in the probability
space of the variables $J_{ij}$) expressing the number of configurations $\eta$ such that
$H(\eta)<-mN$. Denote moreover as $\nu(m,\a)$ the number of configurations $\eta$ such that
$\sum_i\eta_i:=|\eta|=\a N$ and $H(\eta)<-mN$. Note that the possible values of $\a$ are
$\a\in A$ with $A=\{0,\frac{1}{N},\frac{2}{N},\dots,1\}$.
Since the variables $J_{ij}$ are independent, we can write

\begin{align}\label{2.1}
E(\nu(m,\a))={N\choose\a N}\frac{1}{\sqrt{2\pi \alpha^2N}}\int_{-\infty}^{-mN}e^{-\frac{x^2}{2\a^2N}}dx
\end{align}
and therefore

\begin{align}\label{2.2}
E(\nu(m))=\sum_{\a\in A}{N\choose\a N}\frac{1}{\sqrt{2\pi \alpha^2N}}\int_{-\infty}^{-mN}e^{-\frac{x^2}{2\a^2N}}dx
\end{align}
Now we introduce the following notation: given two functions $f(N)$ and $g(N)$ we say that
$f$ and $g$ are \emph{logarithmically equivalent}, and we write $f\asymp g$ if
\begin{align}\label{2.3}
\lim_{N\rightarrow\infty}\frac{1}{N}\log f(N)=\lim_{N\rightarrow\infty}\frac{1}{N}\log g(N)
\end{align}
Using the well known results
\begin{align}\label{2.3.1}
{N\choose\a N}\asymp e^{NI(\a)}
\end{align}
with
$$
I(\a)=-\a\ln\a-(1-\a)\ln(1-\a),
$$
and
$$
\int_{-\infty}^{-mN}e^{-\frac{x^2}{2\a^2N}}dx\asymp e^{-\frac{Nm^2}{2\a^2}}
$$
and the saddle point method, we easily get

\begin{align}\label{2.4}
E(\nu(m))\asymp\max_{\a\in[0,1]}e^{N\left(I(\a)-\frac{Nm^2}{2\a^2}\right)}:=\max_{\a\in[0,1]}e^{NF(\a,m)}
\end{align} 
Standard numerical evaluations show that for $m>.801$ $F(\a,m)<0\ \forall\ \a$. For
$m=.801$, $F(\bar\a,m)=0$ for $\a =.788$, while $F(\a,m)<0\ \forall\ \a\ne\ .788$. Finally,
there exists a set of value of $\a$ such that $F(\a, m)>0$ for all $m<.801$.
Using that $1-P(n=0)<E(n)$, holding for non negative integer random variables $n$,
the results mentioned above show that for $m>.801$ we have $P(\exists \eta:H(\eta)<-mN)\asymp e^{-c_1(m)N}$
for some $c_1(m)>0$.
However this does not give us indication on the set of values $m$ such that 
$P(\exists \eta:H(\eta)<-mN)$
is exponentially close to 1. As mentioned in the introduction,
numerical computations, that are exact evaluations for $N\le150$ and
heuristic estimates for larger $N$, see Section~\ref{sec:a_pca_based_algorithm_to_find_the_minima_of_h} below, show that the critical value of $m$ such that it starts to
appear some configuration with $H(\eta)<-mN$ is around the value $\bar m=.42$, and such configurations
have typically $\a=\bar\a=.62$. Hence, our lower bound of the minimum energy of the system is quite rough.

The reason of this inaccuracy is due to the fact that, although the computation of $E(\nu(m))$
is an exact evaluation, we can use it as a rough estimate in the direct evaluation of the quantity
$$P_0(m):=P(\nu(m)=0)$$
Indeed
\begin{align}\label{2.5}
P_0(m)=1-P(\cup_\eta \{H(\eta)<-mN\})\ge 1-\sum_\eta P(H(\eta)<-mN)
\end{align}
and
$\sum_\eta P(H(\eta)<-mN)$ is exactly the quantity we evaluated above. Here it is clear that
using  $E(\nu(m))$ in order to evaluate $P_0(m)$ we are pretending that the probability of a
union of events (that are far to be disjoint!) is equal to the sum of the probabilities of the
events. This is indeed the reason why the estimate is quite rough. 
Unfortunately it does not seem that the use of
Bonferroni-kind inequalities is of some help.

It is not difficult, however, to improve the estimate above by taking into account the
correlation between the $J_{ij}$ selected by $\eta$ and the rest of them.
Let us call $\bf 1$ the configuration $\eta$ such that $\eta_i=1\ \forall i$.
Clearly $H({\bf 1})=\frac{1}{\sqrt N}\sum_{i,j}J_{ij}$.
We can write, for fixed values of $m,\a$ and for $0<\g<\frac{1}{2}$
$$P\left(\nu(m,\a)=0, |H({\bf 1})|\le N^{\frac{1}{2}+\g}\right)=$$
$$
1-P(|H({\bf 1})|> N^{\frac{1}{2}+\g})-P\left(\bigcup_{|\eta|=\a N}
H(\eta)<-mN,|H({\bf 1})|\le N^{\frac{1}{2}+\g}\right)
$$
Hence we have
$$
P\left(\nu(m,\a)=0\right)\ge P\left(\nu(m,\a)=0, |H({\bf 1})|\le N^{\frac{1}{2}+\g}\right)\ge
$$
$$
\ge1-P(|H({\bf 1})|> N^{\frac{1}{2}+\g})-P\left(\bigcup_{|\eta|=\a N} H(\eta)<-mN,|H({\bf 1})|\le N^{\frac{1}{2}+\g}\right)\ge
$$
$$
\ge1-P(|H({\bf 1})|> N^{\frac{1}{2}+\g})-\sum_{|\eta|=\a N} P\left( H(\eta)<-mN,|H({\bf 1})|\le N^{\frac{1}{2}+\g}\right)=
$$
\begin{align}\label{2.6}=
1-P(|H({\bf 1})|> N^{\frac{1}{2}+\g})-{N\choose\a N}P\left( H(\eta_\a)<-mN,|H({\bf 1})|\le N^{\frac{1}{2}+\g}\right)
\end{align}

Where $\eta_\alpha$ is the configuration such that $\eta_{\a,i}=1$ for $i=1,...,\a N$ and $\eta_{\a,i}=0$ otherwise. Here we use the fact that
$P\left( H(\eta)<-mN,|H({\bf 1})|\le N^{\frac{1}{2}+\g}\right)$ does not depend on the choice
of $\eta$, but only on its cardinality $|\eta|$.
We then evaluate
\begin{align}\label{2.7}
P(|H({\bf 1})|> N^{\frac{1}{2}+\g})=2\frac{1}{\sqrt{2\pi N}}\int_{N^{\frac{1}{2}+\g}}^\infty e^{-\frac{x^2}{2N}}dx\asymp
0
\end{align}
and
$$
P\left( H(\eta)<-mN,|H({\bf 1})|\le N^{\frac{1}{2}+\g}\right)=$$
$$=\frac{1}{\sqrt{2\pi \a^2 N}}\frac{1}{\sqrt{2\pi (1-\a^2)N}}
\int_{-\infty}^{-mN}dx\int_{-N^{\frac{1}{2}+\g}-x}^{N^{\frac{1}{2}+\g}-x}dy
\ e^{-\frac{x^2}{2\a^2N}}e^{-\frac{y^2}{2(1-\a^2)N}}\asymp
$$
\begin{align}\label{2.8}
\int_{-\infty}^{-mN}dx\ e^{-\frac{x^2}{2\a^2(1-\a^2)N}+\frac{2xN^{\frac{1}{2}+\g}-N^{1+2\g}}{2(1-\a^2)N}}\asymp e^{-N\frac{m^2}{2\a^2(1-\a^2)}}
\end{align}
putting \eqref{2.7} and \eqref{2.8} in \eqref{2.6} and exploiting \eqref{2.3.1} we obtain
\begin{align}\label{2.9}
P\left(\nu(m,\a)=0\right)\ge1-G(m,\a)
\end{align}
with
\begin{align}\label{2.10}
G(m,\a)\asymp +e^{N\left(I(\a)-\frac{m^2}{2\a^2(1-\a^2)}\right)}
\end{align}
Call $F_1(m,\a):=I(\a)-\frac{m^2}{2\a^2(1-\a^2)}$
Again, standard numerical evaluations show that for $m>\bar m=.562$
$F_1(\a,\bar m)<0\ \forall\ \a$. For
$m=\bar m$, $F_1(\bar\a,\bar m)=0$ for $\a =\bar\a=.644$, while $F_1(\a,\bar m)<0\ \forall\ \a\ne\bar\a$. Finally,
there exists a set of value of $\a$ such that $F_1(\a, m)>0$ for all $m<\bar m$.

We have proved, therefore, that for $.562<m<.801$ the system is in a regime in which
the probability to have some configuration $\eta$ such that $H(\eta)<-mN$ is exponentially small
in $N$, while the average number of configurations $\eta$ such that $H(\eta)<-mN$ is exponentially
large in $N$. 
This circumstance outlines a particular feature of disordered systems: it happens often that in order
to evaluate the average of some quantity with respect to the realization of the disorder, non typical
configurations (in our case, configuration with large $H({\bf 1})$) give the leading contribution to such
averages.
Note that $\bar m=.562$ is un upper bound uniform in $N$ of the
value of $E(m_N)$.

\section{A PCA based algorithm to find the minima of $H$} 
\label{sec:a_pca_based_algorithm_to_find_the_minima_of_h}

In this section we introduce a heuristic algorithm,
referred to as PCA ($\cav$),
to find the minima of the Hamiltonian $H(\eta)$ defined in \eqref{eq:hamiltonian} and,
more generally, to deal with the Unconstrained Binary Quadratic
Programming whose
objective is to determine
\begin{align}\label{eq:ubqp}
	\min_{\eta} \eta^\text{T}J\eta
\end{align}
where $\eta \in \{0,1\}^N$ and $J$ is a constant $N \times N$ real matrix.

Note that, if instead of J, we consider the symmetric matrix $J' = \frac{J + J^T}{2}$,
the value of \eqref{eq:ubqp}
(and of \eqref{eq:hamiltonian}) does not change.

Our algorithm consists in a Probabilistic Cellular Automaton
defined on the space $\{0,1\}^N$, that is a Markov Chain
$({X_n})_{n\in\N}$ whose transition probabilities are such that
$P\{X_n = \tau | X_{n-1} = \eta\} =
\prod_{i=1} P\{(X_n)_{i} = \tau_{i} | X_{n-1} = \eta\}$.
In other words, a PCA is a Markov Chain for which,
at each step, each component of the ``configuration''
is updated independently from all the others.
From a computational point of view, a Markov Chain
as such is particularly interesting since the simulation
of its evolution is well adapted to be executed on
(massively) parallel processors (for instance GPUs).

If the Markov Chain we defined had an equilibrium distribution
concentrated on those configurations where the minimal values
of the Hamiltonian function $H$ defined in \eqref{eq:hamiltonian}
are attained, then letting the system evolve for a sufficiently
long time would result in a high probability of reaching
some configuration $\eta$ for which $H(\eta)$
is close to $H(\eta^{\star})$ where $\eta^{\star}$ is
the minimizer of $H$.

For this purpose, we introduce the following (Hamiltonian)
function, defined on pairs of configurations,
\begin{align}\label{eq:pairs_hamiltonian}
	H(\eta, \tau) =
		\beta \sum_{i} h_{i}(\eta)\tau_{i}
		+ q \sum_{i} \left[\eta_{i}(1 - \tau_{i}) + \tau_{i} (1 - \eta_{i})\right]
\end{align}
where $h_{i}(\eta) = \frac{1}{\sqrt{N}}\sum_{j} J_{ij} \eta_{j}$, $\beta$
is the inverse temperature, $q$ is a positive constant representing
an inertial term and
$J = \frac{L + L^{T}}{2}$ with $L_{i,j}$
Gaussian random variables with mean $0$ and variance $1$.
We define the (time homogeneous) transition matrix as
\begin{align}\label{eq:config_transition_probabilities}
	P_{\eta, \tau} =
	\frac{e^{-H(\eta, \tau)}}
	{\sum_{\tau}e^{-H(\eta, \tau)}}
\end{align}
and let the system evolve according to
\eqref{eq:config_transition_probabilities}.

Note that $H(\eta, \eta) = \beta H(\eta)$.
Further note that, because of the simmetry of $J$, the transition probabilities defined in
\eqref{eq:config_transition_probabilities}
satisfy
\begin{align}\label{eq:pca_detailed_balance}
	\frac{\sum_{\tau} e^{-H(\eta, \tau)}}
	{\sum_{\eta, \tau} e^{-H(\eta, \tau)}}
	P_{\eta, \tau}
	=
	P_{\tau, \eta}
	\frac{{\sum_{\eta} e^{-H(\eta, \tau)}}}
	{\sum_{\eta, \tau} e^{-H(\eta, \tau)}}
\end{align}
and, therefore,
\begin{align}\label{eq:pca_equilibrium_measure}
	\pi(\eta) = \frac{\sum_{\tau} e^{-H(\eta, \tau)}}
	{\sum_{\eta, \tau} e^{-H(\eta, \tau)}}
\end{align}
is the (reversible) equilibrium measure of the Markov Chain.

The transition matrix \eqref{eq:config_transition_probabilities} can be rewritten in the form
\begin{align}\label{eq:config_transition_probabilities_as_prod}
	P_{\eta, \tau} =
	\prod_{i}
	\frac{
	e^{-\beta h_{i}(\eta)\tau_{i}
	-q\left[\eta_{i} (1-\tau_{i}) + \tau_{i} (1-\eta_{i}) \right]}}
	{Z_{i}}
\end{align}
for appropriate normalization constants
$(Z_{i})_{i = 1, \ldots, N}$,
from which it is immedate to see that, given $\eta$, the probability
that $\tau_{i}$ takes value $0$ or $1$ does not depend on $\tau_{j}$ for all pairs $(i,j)$ and, therefore,
our Markov Chain is indeed a PCA.

In particular, for $i = 1, \ldots, N$ we have
\begin{align}
	P(\tau_{i} = 1| \eta) =
	\frac{e^{-\beta h_{i}(\eta) - q\eta_{i}}}
	{Z_{i}}
\end{align}
and
\begin{align}
	P(\tau_{i} = 0| \eta) =
	\frac{e^{-q(1-\eta_{i})}}
	{Z_{i}}
\end{align}
where $Z_{i} = e^{-\beta h_{i} + q\eta_{i}} + e^{q(1-\eta_{i})}$.

As a consequence, the probability $P(\tau_{i} = \cdot|\eta)$
can be computed independently from $P(\tau_{j} = \cdot|\eta)$
for all $i,j$ and, in principle, each of this probabilities
could be effectively evaluated
on a dedicated processor core.

Unfortunately, we do not have a detailed knowledge of the
probability distribution in \eqref{eq:pca_equilibrium_measure}.
However, the rationale behind the introduction of this algorithm relies
on the observation that,
if $q$ is large, the weight of the terms where
$\eta \neq \tau$ only give a small contribution to the
probability measure \eqref{eq:pca_equilibrium_measure} and, thus, it is
reasonable to expect that, for $q$ big enough,
\begin{align}\label{eq:gibbs_measure}
	\pi(\eta)
	=
	\frac{\sum_{\tau} e^{-H(\eta, \tau)}}
	{\sum_{\eta, \tau} e^{-H(\eta, \tau)}}
	\approx
	\frac{e^{-H(\eta, \eta)}}
	{\sum_{\eta} e^{-H(\eta, \eta)}}
	=
	\frac{e^{-\beta H(\eta)}}
	{\sum_{\eta} e^{-\beta H(\eta)}}
	= \pi_{G}(\eta)
\end{align}
that is, the equilibrium measure $\pi(\eta)$, can
be thought to be close to the Gibbs measure $\pi_{G}(\eta)$ which,
at low temperatures (i.e., when $\beta$ is large),
is concentrated on those configurations where the minimal
values of $H$ are attained.
Therefore, a proper choice of the parameters $\beta$ and $q$ should
let our PCA visit, with high probability, the minimizers of $H$.

In \cite{DSS}, the argument presented above
has been made precise and, in particular, it has been proven that
in the high temperature regime (that is, when $\beta$ is small)
if the thermodynamics limit $N \rightarrow \infty$ is considered,
the measure \eqref{eq:pca_equilibrium_measure} tends to
the Gibbs measure as $q$ goes to $\infty$.
Moreover, in \cite{PSS}, it has been proven that,
for a certain class of models, the same claim
holds for every value of $\beta$.

As an additional motivation behind the introduction
of a PCA to tackle the problem of finding the minima
of a quadratic form in $\{0,1\}^N$, is the fact that
the implementation of a collective dynamics has already been
proven to be very effective to face a related problem, namely
the Clique Problem \cite{ISS}, \cite{GSSV}.

\subsection{Behavior of the minima of $H$}
We used the PCA algorithm described above
to collect some statistics on $m_{N}$
and  the proportion of $1$'s in the configuration of minimal energy when the
system size is $N$. In what follows we denote this quantity by $\alpha_{N}$.
Namely, we estimated
$E(m_{N})$ and $\text{Var}({m_{N}})$ and computed the average of $\alpha_{N}$
for values of $N$ ranging from $1$ to $500$.

In order to obtain some confidence on our estimates, we
performed multiple independent runs of our algorithm on each
randomly generated instance and kept only those results for which the
multiple runs found the same minimizer.

Clearly, this procedure does not guarantee that the algorithm has
reached a minimizer of $H$, nevertheless we expect the statistics collected
in this way to be quite reliable.

For instances of size up to $20$ the minima are found by direct inspection.

For instances of size $N \le 150$ we were able to compare,
for some of the tested instances, the heuristic minima obtained with
our algorithm with the exact minima obtained with the
SDP based Branch \& Bound algorithm
developed by Rendl, Rinaldi and Wiegele \cite{RRW10} and we found
that the heuristic minima coincided with the exact ones.

\begin{figure}
	\centering
	\resizebox{.9\textwidth}{!}{\input{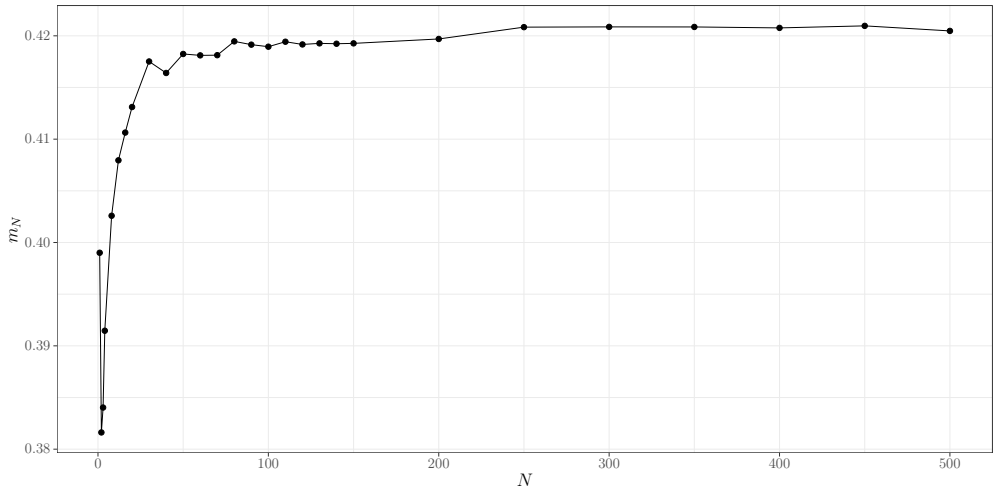}}
	\caption{The estimated $m_{N}$ as a function of $N$}
	\label{fig_min_as_N_grows}
\end{figure}

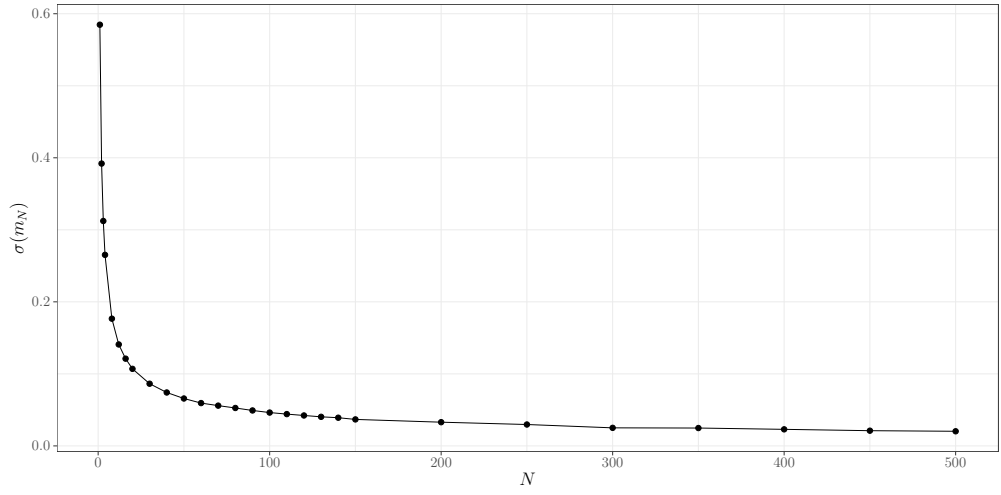
\begin{figure}
	\centering
	\resizebox{.9\textwidth}{!}{
\begin{tikzpicture}[x=1pt,y=1pt]
\definecolor{fillColor}{RGB}{255,255,255}
\path[use as bounding box,fill=fillColor,fill opacity=0.00] (0,0) rectangle (722.70,361.35);
\begin{scope}
\path[clip] (  0.00,  0.00) rectangle (722.70,361.35);
\definecolor{drawColor}{RGB}{255,255,255}
\definecolor{fillColor}{RGB}{255,255,255}

\path[draw=drawColor,line width= 0.6pt,line join=round,line cap=round,fill=fillColor] (  0.00, -0.00) rectangle (722.70,361.35);
\end{scope}
\begin{scope}
\path[clip] ( 37.27, 30.14) rectangle (717.20,355.85);
\definecolor{fillColor}{RGB}{255,255,255}

\path[fill=fillColor] ( 37.27, 30.14) rectangle (717.20,355.85);
\definecolor{drawColor}{gray}{0.92}

\path[draw=drawColor,line width= 0.3pt,line join=round] ( 37.27, 86.76) --
	(717.20, 86.76);

\path[draw=drawColor,line width= 0.3pt,line join=round] ( 37.27,191.67) --
	(717.20,191.67);

\path[draw=drawColor,line width= 0.3pt,line join=round] ( 37.27,296.59) --
	(717.20,296.59);

\path[draw=drawColor,line width= 0.3pt,line join=round] (128.87, 30.14) --
	(128.87,355.85);

\path[draw=drawColor,line width= 0.3pt,line join=round] (252.74, 30.14) --
	(252.74,355.85);

\path[draw=drawColor,line width= 0.3pt,line join=round] (376.61, 30.14) --
	(376.61,355.85);

\path[draw=drawColor,line width= 0.3pt,line join=round] (500.49, 30.14) --
	(500.49,355.85);

\path[draw=drawColor,line width= 0.3pt,line join=round] (624.36, 30.14) --
	(624.36,355.85);

\path[draw=drawColor,line width= 0.6pt,line join=round] ( 37.27, 34.30) --
	(717.20, 34.30);

\path[draw=drawColor,line width= 0.6pt,line join=round] ( 37.27,139.22) --
	(717.20,139.22);

\path[draw=drawColor,line width= 0.6pt,line join=round] ( 37.27,244.13) --
	(717.20,244.13);

\path[draw=drawColor,line width= 0.6pt,line join=round] ( 37.27,349.04) --
	(717.20,349.04);

\path[draw=drawColor,line width= 0.6pt,line join=round] ( 66.93, 30.14) --
	( 66.93,355.85);

\path[draw=drawColor,line width= 0.6pt,line join=round] (190.81, 30.14) --
	(190.81,355.85);

\path[draw=drawColor,line width= 0.6pt,line join=round] (314.68, 30.14) --
	(314.68,355.85);

\path[draw=drawColor,line width= 0.6pt,line join=round] (438.55, 30.14) --
	(438.55,355.85);

\path[draw=drawColor,line width= 0.6pt,line join=round] (562.42, 30.14) --
	(562.42,355.85);

\path[draw=drawColor,line width= 0.6pt,line join=round] (686.29, 30.14) --
	(686.29,355.85);
\definecolor{drawColor}{RGB}{0,0,0}

\path[draw=drawColor,line width= 0.6pt,line join=round] ( 68.17,341.05) --
	( 69.41,239.92) --
	( 70.65,198.10) --
	( 71.89,173.43) --
	( 76.84,126.96) --
	( 81.80,108.18) --
	( 86.75, 97.84) --
	( 91.71, 90.42) --
	(104.10, 79.59) --
	(116.48, 73.20) --
	(128.87, 68.82) --
	(141.26, 65.49) --
	(153.64, 63.63) --
	(166.03, 61.92) --
	(178.42, 60.11) --
	(190.81, 58.58) --
	(203.19, 57.46) --
	(215.58, 56.46) --
	(227.97, 55.48) --
	(240.36, 54.83) --
	(252.74, 53.59) --
	(314.68, 51.57) --
	(376.61, 49.88) --
	(438.55, 47.45) --
	(500.49, 47.31) --
	(562.42, 46.36) --
	(624.36, 45.38) --
	(686.29, 44.95);
\definecolor{fillColor}{RGB}{0,0,0}

\path[draw=drawColor,line width= 0.4pt,line join=round,line cap=round,fill=fillColor] ( 68.17,341.05) circle (  1.96);

\path[draw=drawColor,line width= 0.4pt,line join=round,line cap=round,fill=fillColor] ( 69.41,239.92) circle (  1.96);

\path[draw=drawColor,line width= 0.4pt,line join=round,line cap=round,fill=fillColor] ( 70.65,198.10) circle (  1.96);

\path[draw=drawColor,line width= 0.4pt,line join=round,line cap=round,fill=fillColor] ( 71.89,173.43) circle (  1.96);

\path[draw=drawColor,line width= 0.4pt,line join=round,line cap=round,fill=fillColor] ( 76.84,126.96) circle (  1.96);

\path[draw=drawColor,line width= 0.4pt,line join=round,line cap=round,fill=fillColor] ( 81.80,108.18) circle (  1.96);

\path[draw=drawColor,line width= 0.4pt,line join=round,line cap=round,fill=fillColor] ( 86.75, 97.84) circle (  1.96);

\path[draw=drawColor,line width= 0.4pt,line join=round,line cap=round,fill=fillColor] ( 91.71, 90.42) circle (  1.96);

\path[draw=drawColor,line width= 0.4pt,line join=round,line cap=round,fill=fillColor] (104.10, 79.59) circle (  1.96);

\path[draw=drawColor,line width= 0.4pt,line join=round,line cap=round,fill=fillColor] (116.48, 73.20) circle (  1.96);

\path[draw=drawColor,line width= 0.4pt,line join=round,line cap=round,fill=fillColor] (128.87, 68.82) circle (  1.96);

\path[draw=drawColor,line width= 0.4pt,line join=round,line cap=round,fill=fillColor] (141.26, 65.49) circle (  1.96);

\path[draw=drawColor,line width= 0.4pt,line join=round,line cap=round,fill=fillColor] (153.64, 63.63) circle (  1.96);

\path[draw=drawColor,line width= 0.4pt,line join=round,line cap=round,fill=fillColor] (166.03, 61.92) circle (  1.96);

\path[draw=drawColor,line width= 0.4pt,line join=round,line cap=round,fill=fillColor] (178.42, 60.11) circle (  1.96);

\path[draw=drawColor,line width= 0.4pt,line join=round,line cap=round,fill=fillColor] (190.81, 58.58) circle (  1.96);

\path[draw=drawColor,line width= 0.4pt,line join=round,line cap=round,fill=fillColor] (203.19, 57.46) circle (  1.96);

\path[draw=drawColor,line width= 0.4pt,line join=round,line cap=round,fill=fillColor] (215.58, 56.46) circle (  1.96);

\path[draw=drawColor,line width= 0.4pt,line join=round,line cap=round,fill=fillColor] (227.97, 55.48) circle (  1.96);

\path[draw=drawColor,line width= 0.4pt,line join=round,line cap=round,fill=fillColor] (240.36, 54.83) circle (  1.96);

\path[draw=drawColor,line width= 0.4pt,line join=round,line cap=round,fill=fillColor] (252.74, 53.59) circle (  1.96);

\path[draw=drawColor,line width= 0.4pt,line join=round,line cap=round,fill=fillColor] (314.68, 51.57) circle (  1.96);

\path[draw=drawColor,line width= 0.4pt,line join=round,line cap=round,fill=fillColor] (376.61, 49.88) circle (  1.96);

\path[draw=drawColor,line width= 0.4pt,line join=round,line cap=round,fill=fillColor] (438.55, 47.45) circle (  1.96);

\path[draw=drawColor,line width= 0.4pt,line join=round,line cap=round,fill=fillColor] (500.49, 47.31) circle (  1.96);

\path[draw=drawColor,line width= 0.4pt,line join=round,line cap=round,fill=fillColor] (562.42, 46.36) circle (  1.96);

\path[draw=drawColor,line width= 0.4pt,line join=round,line cap=round,fill=fillColor] (624.36, 45.38) circle (  1.96);

\path[draw=drawColor,line width= 0.4pt,line join=round,line cap=round,fill=fillColor] (686.29, 44.95) circle (  1.96);
\definecolor{drawColor}{gray}{0.20}

\path[draw=drawColor,line width= 0.6pt,line join=round,line cap=round] ( 37.27, 30.14) rectangle (717.20,355.85);
\end{scope}
\begin{scope}
\path[clip] (  0.00,  0.00) rectangle (722.70,361.35);
\definecolor{drawColor}{gray}{0.30}

\node[text=drawColor,anchor=base east,inner sep=0pt, outer sep=0pt, scale=  0.88] at ( 32.32, 31.27) {0.0};

\node[text=drawColor,anchor=base east,inner sep=0pt, outer sep=0pt, scale=  0.88] at ( 32.32,136.19) {0.2};

\node[text=drawColor,anchor=base east,inner sep=0pt, outer sep=0pt, scale=  0.88] at ( 32.32,241.10) {0.4};

\node[text=drawColor,anchor=base east,inner sep=0pt, outer sep=0pt, scale=  0.88] at ( 32.32,346.01) {0.6};
\end{scope}
\begin{scope}
\path[clip] (  0.00,  0.00) rectangle (722.70,361.35);
\definecolor{drawColor}{gray}{0.20}

\path[draw=drawColor,line width= 0.6pt,line join=round] ( 34.52, 34.30) --
	( 37.27, 34.30);

\path[draw=drawColor,line width= 0.6pt,line join=round] ( 34.52,139.22) --
	( 37.27,139.22);

\path[draw=drawColor,line width= 0.6pt,line join=round] ( 34.52,244.13) --
	( 37.27,244.13);

\path[draw=drawColor,line width= 0.6pt,line join=round] ( 34.52,349.04) --
	( 37.27,349.04);
\end{scope}
\begin{scope}
\path[clip] (  0.00,  0.00) rectangle (722.70,361.35);
\definecolor{drawColor}{gray}{0.20}

\path[draw=drawColor,line width= 0.6pt,line join=round] ( 66.93, 27.39) --
	( 66.93, 30.14);

\path[draw=drawColor,line width= 0.6pt,line join=round] (190.81, 27.39) --
	(190.81, 30.14);

\path[draw=drawColor,line width= 0.6pt,line join=round] (314.68, 27.39) --
	(314.68, 30.14);

\path[draw=drawColor,line width= 0.6pt,line join=round] (438.55, 27.39) --
	(438.55, 30.14);

\path[draw=drawColor,line width= 0.6pt,line join=round] (562.42, 27.39) --
	(562.42, 30.14);

\path[draw=drawColor,line width= 0.6pt,line join=round] (686.29, 27.39) --
	(686.29, 30.14);
\end{scope}
\begin{scope}
\path[clip] (  0.00,  0.00) rectangle (722.70,361.35);
\definecolor{drawColor}{gray}{0.30}

\node[text=drawColor,anchor=base,inner sep=0pt, outer sep=0pt, scale=  0.88] at ( 66.93, 19.13) {0};

\node[text=drawColor,anchor=base,inner sep=0pt, outer sep=0pt, scale=  0.88] at (190.81, 19.13) {100};

\node[text=drawColor,anchor=base,inner sep=0pt, outer sep=0pt, scale=  0.88] at (314.68, 19.13) {200};

\node[text=drawColor,anchor=base,inner sep=0pt, outer sep=0pt, scale=  0.88] at (438.55, 19.13) {300};

\node[text=drawColor,anchor=base,inner sep=0pt, outer sep=0pt, scale=  0.88] at (562.42, 19.13) {400};

\node[text=drawColor,anchor=base,inner sep=0pt, outer sep=0pt, scale=  0.88] at (686.29, 19.13) {500};
\end{scope}
\begin{scope}
\path[clip] (  0.00,  0.00) rectangle (722.70,361.35);
\definecolor{drawColor}{RGB}{0,0,0}

\node[text=drawColor,anchor=base,inner sep=0pt, outer sep=0pt, scale=  1.10] at (377.23,  6.06) {$N$};
\end{scope}
\begin{scope}
\path[clip] (  0.00,  0.00) rectangle (722.70,361.35);
\definecolor{drawColor}{RGB}{0,0,0}

\node[text=drawColor,rotate= 90.00,anchor=base,inner sep=0pt, outer sep=0pt, scale=  1.10] at ( 13.08,193.00) {$\sigma(m_{N})$};
\end{scope}
\end{tikzpicture}}
	\caption{The estimated standard deviation of $m_{N}$ as a function of $N$}
	\label{fig_var_min_as_N_grows}
\end{figure}

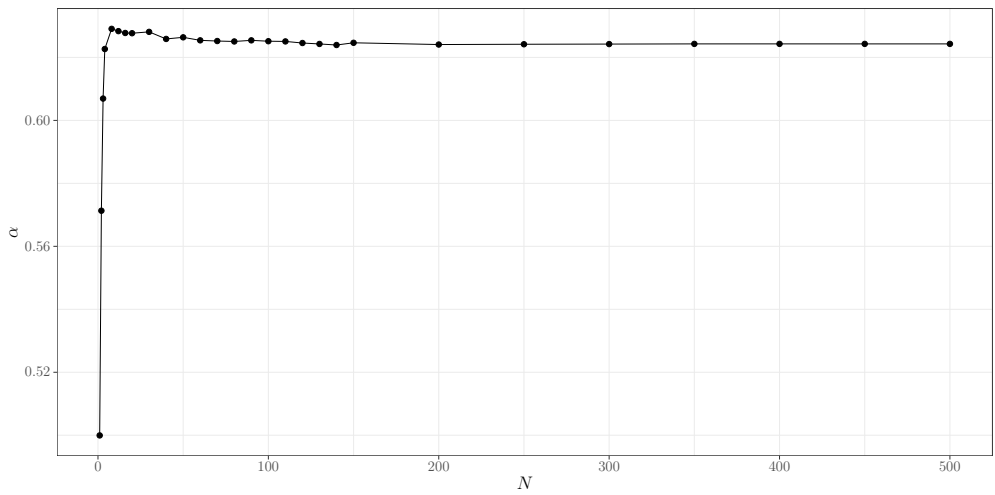
\begin{figure}
	\centering
	\resizebox{.9\textwidth}{!}{
\begin{tikzpicture}[x=1pt,y=1pt]
\definecolor{fillColor}{RGB}{255,255,255}
\path[use as bounding box,fill=fillColor,fill opacity=0.00] (0,0) rectangle (722.70,361.35);
\begin{scope}
\path[clip] (  0.00,  0.00) rectangle (722.70,361.35);
\definecolor{drawColor}{RGB}{255,255,255}
\definecolor{fillColor}{RGB}{255,255,255}

\path[draw=drawColor,line width= 0.6pt,line join=round,line cap=round,fill=fillColor] (  0.00, -0.00) rectangle (722.70,361.35);
\end{scope}
\begin{scope}
\path[clip] ( 41.67, 30.14) rectangle (717.20,355.85);
\definecolor{fillColor}{RGB}{255,255,255}

\path[fill=fillColor] ( 41.67, 30.14) rectangle (717.20,355.85);
\definecolor{drawColor}{gray}{0.92}

\path[draw=drawColor,line width= 0.3pt,line join=round] ( 41.67, 45.08) --
	(717.20, 45.08);

\path[draw=drawColor,line width= 0.3pt,line join=round] ( 41.67,136.80) --
	(717.20,136.80);

\path[draw=drawColor,line width= 0.3pt,line join=round] ( 41.67,228.51) --
	(717.20,228.51);

\path[draw=drawColor,line width= 0.3pt,line join=round] ( 41.67,320.23) --
	(717.20,320.23);

\path[draw=drawColor,line width= 0.3pt,line join=round] (132.68, 30.14) --
	(132.68,355.85);

\path[draw=drawColor,line width= 0.3pt,line join=round] (255.75, 30.14) --
	(255.75,355.85);

\path[draw=drawColor,line width= 0.3pt,line join=round] (378.82, 30.14) --
	(378.82,355.85);

\path[draw=drawColor,line width= 0.3pt,line join=round] (501.89, 30.14) --
	(501.89,355.85);

\path[draw=drawColor,line width= 0.3pt,line join=round] (624.96, 30.14) --
	(624.96,355.85);

\path[draw=drawColor,line width= 0.6pt,line join=round] ( 41.67, 90.94) --
	(717.20, 90.94);

\path[draw=drawColor,line width= 0.6pt,line join=round] ( 41.67,182.66) --
	(717.20,182.66);

\path[draw=drawColor,line width= 0.6pt,line join=round] ( 41.67,274.37) --
	(717.20,274.37);

\path[draw=drawColor,line width= 0.6pt,line join=round] ( 71.14, 30.14) --
	( 71.14,355.85);

\path[draw=drawColor,line width= 0.6pt,line join=round] (194.21, 30.14) --
	(194.21,355.85);

\path[draw=drawColor,line width= 0.6pt,line join=round] (317.28, 30.14) --
	(317.28,355.85);

\path[draw=drawColor,line width= 0.6pt,line join=round] (440.35, 30.14) --
	(440.35,355.85);

\path[draw=drawColor,line width= 0.6pt,line join=round] (563.42, 30.14) --
	(563.42,355.85);

\path[draw=drawColor,line width= 0.6pt,line join=round] (686.49, 30.14) --
	(686.49,355.85);
\definecolor{drawColor}{RGB}{0,0,0}

\path[draw=drawColor,line width= 0.6pt,line join=round] ( 72.37, 44.95) --
	( 73.60,208.59) --
	( 74.83,290.28) --
	( 76.06,326.36) --
	( 80.99,341.05) --
	( 85.91,339.38) --
	( 90.83,338.04) --
	( 95.76,337.86) --
	(108.06,338.85) --
	(120.37,333.70) --
	(132.68,334.87) --
	(144.98,332.67) --
	(157.29,332.18) --
	(169.60,331.86) --
	(181.90,332.67) --
	(194.21,332.04) --
	(206.52,331.89) --
	(218.83,330.73) --
	(231.13,330.05) --
	(243.44,329.28) --
	(255.75,330.89) --
	(317.28,329.62) --
	(378.82,329.83) --
	(440.35,329.96) --
	(501.89,330.09) --
	(563.42,330.10) --
	(624.96,330.09) --
	(686.49,330.09);
\definecolor{fillColor}{RGB}{0,0,0}

\path[draw=drawColor,line width= 0.4pt,line join=round,line cap=round,fill=fillColor] ( 72.37, 44.95) circle (  1.96);

\path[draw=drawColor,line width= 0.4pt,line join=round,line cap=round,fill=fillColor] ( 73.60,208.59) circle (  1.96);

\path[draw=drawColor,line width= 0.4pt,line join=round,line cap=round,fill=fillColor] ( 74.83,290.28) circle (  1.96);

\path[draw=drawColor,line width= 0.4pt,line join=round,line cap=round,fill=fillColor] ( 76.06,326.36) circle (  1.96);

\path[draw=drawColor,line width= 0.4pt,line join=round,line cap=round,fill=fillColor] ( 80.99,341.05) circle (  1.96);

\path[draw=drawColor,line width= 0.4pt,line join=round,line cap=round,fill=fillColor] ( 85.91,339.38) circle (  1.96);

\path[draw=drawColor,line width= 0.4pt,line join=round,line cap=round,fill=fillColor] ( 90.83,338.04) circle (  1.96);

\path[draw=drawColor,line width= 0.4pt,line join=round,line cap=round,fill=fillColor] ( 95.76,337.86) circle (  1.96);

\path[draw=drawColor,line width= 0.4pt,line join=round,line cap=round,fill=fillColor] (108.06,338.85) circle (  1.96);

\path[draw=drawColor,line width= 0.4pt,line join=round,line cap=round,fill=fillColor] (120.37,333.70) circle (  1.96);

\path[draw=drawColor,line width= 0.4pt,line join=round,line cap=round,fill=fillColor] (132.68,334.87) circle (  1.96);

\path[draw=drawColor,line width= 0.4pt,line join=round,line cap=round,fill=fillColor] (144.98,332.67) circle (  1.96);

\path[draw=drawColor,line width= 0.4pt,line join=round,line cap=round,fill=fillColor] (157.29,332.18) circle (  1.96);

\path[draw=drawColor,line width= 0.4pt,line join=round,line cap=round,fill=fillColor] (169.60,331.86) circle (  1.96);

\path[draw=drawColor,line width= 0.4pt,line join=round,line cap=round,fill=fillColor] (181.90,332.67) circle (  1.96);

\path[draw=drawColor,line width= 0.4pt,line join=round,line cap=round,fill=fillColor] (194.21,332.04) circle (  1.96);

\path[draw=drawColor,line width= 0.4pt,line join=round,line cap=round,fill=fillColor] (206.52,331.89) circle (  1.96);

\path[draw=drawColor,line width= 0.4pt,line join=round,line cap=round,fill=fillColor] (218.83,330.73) circle (  1.96);

\path[draw=drawColor,line width= 0.4pt,line join=round,line cap=round,fill=fillColor] (231.13,330.05) circle (  1.96);

\path[draw=drawColor,line width= 0.4pt,line join=round,line cap=round,fill=fillColor] (243.44,329.28) circle (  1.96);

\path[draw=drawColor,line width= 0.4pt,line join=round,line cap=round,fill=fillColor] (255.75,330.89) circle (  1.96);

\path[draw=drawColor,line width= 0.4pt,line join=round,line cap=round,fill=fillColor] (317.28,329.62) circle (  1.96);

\path[draw=drawColor,line width= 0.4pt,line join=round,line cap=round,fill=fillColor] (378.82,329.83) circle (  1.96);

\path[draw=drawColor,line width= 0.4pt,line join=round,line cap=round,fill=fillColor] (440.35,329.96) circle (  1.96);

\path[draw=drawColor,line width= 0.4pt,line join=round,line cap=round,fill=fillColor] (501.89,330.09) circle (  1.96);

\path[draw=drawColor,line width= 0.4pt,line join=round,line cap=round,fill=fillColor] (563.42,330.10) circle (  1.96);

\path[draw=drawColor,line width= 0.4pt,line join=round,line cap=round,fill=fillColor] (624.96,330.09) circle (  1.96);

\path[draw=drawColor,line width= 0.4pt,line join=round,line cap=round,fill=fillColor] (686.49,330.09) circle (  1.96);
\definecolor{drawColor}{gray}{0.20}

\path[draw=drawColor,line width= 0.6pt,line join=round,line cap=round] ( 41.67, 30.14) rectangle (717.20,355.85);
\end{scope}
\begin{scope}
\path[clip] (  0.00,  0.00) rectangle (722.70,361.35);
\definecolor{drawColor}{gray}{0.30}

\node[text=drawColor,anchor=base east,inner sep=0pt, outer sep=0pt, scale=  0.88] at ( 36.72, 87.91) {0.52};

\node[text=drawColor,anchor=base east,inner sep=0pt, outer sep=0pt, scale=  0.88] at ( 36.72,179.62) {0.56};

\node[text=drawColor,anchor=base east,inner sep=0pt, outer sep=0pt, scale=  0.88] at ( 36.72,271.34) {0.60};
\end{scope}
\begin{scope}
\path[clip] (  0.00,  0.00) rectangle (722.70,361.35);
\definecolor{drawColor}{gray}{0.20}

\path[draw=drawColor,line width= 0.6pt,line join=round] ( 38.92, 90.94) --
	( 41.67, 90.94);

\path[draw=drawColor,line width= 0.6pt,line join=round] ( 38.92,182.66) --
	( 41.67,182.66);

\path[draw=drawColor,line width= 0.6pt,line join=round] ( 38.92,274.37) --
	( 41.67,274.37);
\end{scope}
\begin{scope}
\path[clip] (  0.00,  0.00) rectangle (722.70,361.35);
\definecolor{drawColor}{gray}{0.20}

\path[draw=drawColor,line width= 0.6pt,line join=round] ( 71.14, 27.39) --
	( 71.14, 30.14);

\path[draw=drawColor,line width= 0.6pt,line join=round] (194.21, 27.39) --
	(194.21, 30.14);

\path[draw=drawColor,line width= 0.6pt,line join=round] (317.28, 27.39) --
	(317.28, 30.14);

\path[draw=drawColor,line width= 0.6pt,line join=round] (440.35, 27.39) --
	(440.35, 30.14);

\path[draw=drawColor,line width= 0.6pt,line join=round] (563.42, 27.39) --
	(563.42, 30.14);

\path[draw=drawColor,line width= 0.6pt,line join=round] (686.49, 27.39) --
	(686.49, 30.14);
\end{scope}
\begin{scope}
\path[clip] (  0.00,  0.00) rectangle (722.70,361.35);
\definecolor{drawColor}{gray}{0.30}

\node[text=drawColor,anchor=base,inner sep=0pt, outer sep=0pt, scale=  0.88] at ( 71.14, 19.13) {0};

\node[text=drawColor,anchor=base,inner sep=0pt, outer sep=0pt, scale=  0.88] at (194.21, 19.13) {100};

\node[text=drawColor,anchor=base,inner sep=0pt, outer sep=0pt, scale=  0.88] at (317.28, 19.13) {200};

\node[text=drawColor,anchor=base,inner sep=0pt, outer sep=0pt, scale=  0.88] at (440.35, 19.13) {300};

\node[text=drawColor,anchor=base,inner sep=0pt, outer sep=0pt, scale=  0.88] at (563.42, 19.13) {400};

\node[text=drawColor,anchor=base,inner sep=0pt, outer sep=0pt, scale=  0.88] at (686.49, 19.13) {500};
\end{scope}
\begin{scope}
\path[clip] (  0.00,  0.00) rectangle (722.70,361.35);
\definecolor{drawColor}{RGB}{0,0,0}

\node[text=drawColor,anchor=base,inner sep=0pt, outer sep=0pt, scale=  1.10] at (379.43,  6.06) {$N$};
\end{scope}
\begin{scope}
\path[clip] (  0.00,  0.00) rectangle (722.70,361.35);
\definecolor{drawColor}{RGB}{0,0,0}

\node[text=drawColor,rotate= 90.00,anchor=base,inner sep=0pt, outer sep=0pt, scale=  1.10] at ( 13.08,193.00) {$\alpha$};
\end{scope}
\end{tikzpicture}}
	\caption{The estimated $\alpha_{N}$ as a function of $N$}
	\label{fig_alpha_as_N_grows}
\end{figure}

The results obtained are summarized in Table~\ref{tab_minima_as_N_grows}
and in Figures~\ref{fig_min_as_N_grows}, \ref{fig_var_min_as_N_grows}
and \ref{fig_alpha_as_N_grows}.

Our simulations show quite clearly that $m_{N}$ and $\alpha_{N}$
settle very rapidly to values around $.42$ and $.62$ respectively.
Therefore, we expect that $m_{N}$ and $\alpha_{N}$ actually have limits
$m$ and $\alpha$ and that these limits are indeed close to $.42$ and
$.62$.

From the simulations it is also possible to see that the variance of $m_{N}$
decreases quite quickly in $N$ in accordance with the findings of
theorem \ref{thm_small_fluctuations}.

\begin{table}[]
\centering
\begin{tabular}{@{}rlll@{}}
\toprule
\multicolumn{1}{c}{$N$} & \multicolumn{1}{c}{$m_{N}$} & \multicolumn{1}{c}{$\text{Var}(m_{N})$} & \multicolumn{1}{c}{$\alpha$} \\ \midrule
1                       & 0.39900688                 & 0.34193425                              & 0.49994000                  \\
2                       & 0.38161935                 & 0.15364853                              & 0.57131000                  \\
3                       & 0.38402868                 & 0.09750664                              & 0.60694000                  \\
4                       & 0.39146143                 & 0.07034040                              & 0.62267500                  \\
8                       & 0.40258731                 & 0.03120213                              & 0.62908000                  \\
12                      & 0.40795292                 & 0.01983398                              & 0.62835333                  \\
16                      & 0.41063769                 & 0.01467275                              & 0.62776750                  \\
20                      & 0.41311072                 & 0.01144546                              & 0.62769050                  \\
30                      & 0.41751744                 & 0.00745522                              & 0.62812333                  \\
40                      & 0.41640474                 & 0.00549822                              & 0.62587500                  \\
50                      & 0.41824325                 & 0.00432999                              & 0.62638600                  \\
60                      & 0.41810818                 & 0.00353408                              & 0.62542833                  \\
70                      & 0.41812609                 & 0.00312537                              & 0.62521143                  \\
80                      & 0.41946192                 & 0.00277294                              & 0.62507500                  \\
90                      & 0.41914527                 & 0.00242012                              & 0.62542778                  \\
100                     & 0.41894795                 & 0.00214175                              & 0.62515400                  \\
110                     & 0.41942923                 & 0.00194858                              & 0.62508909                  \\
120                     & 0.41916496                 & 0.00178374                              & 0.62458333                  \\
130                     & 0.41926967                 & 0.00162928                              & 0.62428462                  \\
140                     & 0.41923424                 & 0.00153119                              & 0.62395000                  \\
150                     & 0.41926961                 & 0.00135212                              & 0.62465267                  \\
200                     & 0.41969536                 & 0.00108424                              & 0.62409500                  \\
250                     & 0.42083938                 & 0.00088238                              & 0.62418800                  \\
300                     & 0.42086376                 & 0.00062833                              & 0.62424667                  \\
350                     & 0.42085872                 & 0.00061472                              & 0.62430286                  \\
400                     & 0.42076682                 & 0.00052816                              & 0.62430500                  \\
450                     & 0.42096274                 & 0.00044608                              & 0.62430222                  \\
500                     & 0.42047454                 & 0.00041183                              & 0.62430400                  \\ \bottomrule
\end{tabular}
\caption{Estimates of $m_{N}$, $Var(m_{N})$ and $\alpha_{N}$ for values of $N$ between $1$ and $500$}
\label{tab_minima_as_N_grows}
\end{table}

\subsection{Numerical results}
\label{sub:numerical_results}

In this section we present some preliminary numerical results obtained
with the algorithm ($\cav$) introduced above.
We stress that the analysis we performed is far to be exhaustive
and it is mainly aimed at figuring out whether the algorithm deserves to be
studied in further detail.

To assess the performaces of the PCA algorithm ($\cav$),
we compared the results with those obtained
with a simple deterministic greedy algorithm ($\gre$)
and a Metropolis algorithm ($\met$). The latter is a well established
choice to address an optimization problem through a Markov Chain Monte Carlo
(MCMC). Both algorithms are described below.


\subsubsection{A greedy algorithm} 
\label{sub:a_greedy_algorithm}

The algorithm starts by choosing $\eta_{0} = [0, \ldots, 0]$ as initial configuration.
The solution is build by considering a sequence of configurations of increasing cardinality
such that $|\eta_n| = n$ (that is configuration $\eta_{n}$ has $n$ components equal to $1$
and $N-n$ equal to $0$).
Let $\eta_{n}^{i}$ be the configuration obtained from $\eta_{n}$ by setting to $1$ its $i$-component.
Then $\eta_{n+1} = {\argmin}_{i} H(\eta_{n}^{i}) - H(\eta_{n})$.
If $H(\eta_{n}^{i}) < H(\eta_{n})$ the algorithm is allowed a further step, otherwise it returns
$\eta_{n}$ as solution.


\subsubsection{A Metropolis algorithm} 
\label{sub:a_metropolis_algorithm}

This algorithm, as well as the PCA $\cav$,
consists of a Markov Chain defined on the space
$\{0,1\}^{N}$.
State $\eta_{0} = [0, \ldots, 0]$
is chosen as initial configuration
and the system is let to evolve according to
the transition probabilities
\begin{align}\label{eq:metropolis_transition_probabilities}
	P_{\eta, \tau} =
	 \begin{cases}
            	\frac{1}{N}e^{-\beta \left[H(\tau) - H(\eta)\right]_{+}}  & \text{if } \tau = \eta^{i}\\
   		1 - \sum_{i = 1, \ldots, N} \frac{1}{N}e^{-\beta \left[H(\eta^{i}) - H(\eta)\right]_{+}}				& \text{it } \tau = \eta \\
		0       								& \text{otherwise}
        \end{cases}
\end{align}
where $H$ is the Hamiltonian defined in
\eqref{eq:hamiltonian}, $\eta^{i}$ is the configuration obtained from $\eta$
by ``flipping'' ($0 \leftrightarrow 1$) the $i$-th component of $\eta$,  $\beta > 0$ is the inverse temperature
and $\left[a\right]_{+} = \max\{0, a\}$. In words it means that, at each step,
a component of the current configuration is chosen at random and it is ``flipped'' with probability
one if the ``flip'' leads to a configuration of lower energy and it is ``flipped'' with a probability
that is exponentially small if the ``flip'' produces a configuration of higher energy
with the exponent proportional to
the difference of the energy
of the new configuration and the energy of
the current configuration.

The Markov chain defined in
\eqref{eq:metropolis_transition_probabilities}
is reversible with respect to the Gibbs measure
\begin{align}\label{eq:metropolis_equilibrium_measure}
	\pi_G(\eta) = \frac{e^{-\beta H(\eta)}}{Z}.
\end{align}
This can be verified by observing that the
detailed balanced condition\linebreak
$\pi_G(\eta)P_{\eta, \tau} = P_{\tau, \eta}\pi_G(\tau)$
holds.

If $\beta$ is sufficiently large, this measure
is concentrated on the minimizers of $H$ and, hence,
in the long run, the Markov chain
is expected to visit those configurations where the
minima of $H$ are attained.


\subsubsection{Experimental details} 
\label{sub:experimental_details}
We do not have apriori knowledge of the mixing times of the Markov chains underlying
$\cav$ and $\met$ algorithms.
Therefore, the number of iterations of both algorithms
has been established in a somewhat arbitrary manner.
In the attempt of establishing a fair comparison of the two random algorithms we
decided to allow both the same number of ``attempted spin flips'', that is, for a given instance,
the number of times any of the two algorithms tests whether
the value of $\eta_{i}$, for some $i$,
should be updated is the same.
Since at each iteration the PCA algorithm tests
whether each of the $N$ components of the
``current'' configuration should be updated,
for the two algorithms to make the same number
of ``attempted spin flips'' the Metropolis
algorithm should be allowed
$m*N$ iterations where $N$ is the
number of variables of the instance and
$m$ is the number of iterations of the
PCA algorithm for the same instance.

In our experimental setting, we set $m = 10000$
for all the instances taken into account.
Clearly this means that, for larger instances, we
should expect that the Markov chains underlying the
two algorithms are further away from their
equilibrium distributions than they are for
smaller instances.

We lack apriori knowledge also for what concerns the optimal values
of the parameters $\beta$ (for both random algorithms) and $q$
(for $\cav$).
Determining such optimal values appears to be
a challenging task since we do not have a detailed characterization of
the ``energy'' landscape induced by the transition probabilities and,
for the $\cav$ algorithm, because of
the two--dimensional nature
of the space of parameters ($\beta$ and $q$).

In our experiments with larger instances (size from 500 to 8000)
we tested values of $\beta$ (for both algorithms) in the set
$\{0.3, 0.7, 1.1, 1.5, 1.9, 2.3\}$ and values of
$q$ in the set $\{0.5, 1.0, 1.5, 2.0, 2.5\}$
(for $\cav$). Also these values have been chosen without the support
of theoreticl knowledge.
However, both algorithms appeared to be quite robust and gave good results for
different values of the parameters.

All three
algorithms (PCA, Metropolis, and Greedy)
have been implemented in Python. The compiled
NumPy library has been used for the computationally heavier tasks,
namely the random numbers generation
and all linear algbra operations.



\subsubsection{Numerical comparisons}
\label{sub:numerical_comparisons}

We considered several instances of size $500$, $1000$, $2000$, $4000$ and $8000$.
For each instance, several runs of both the $\cav$ and $\met$ algorithms have been
executed and the best result for each algorithm have been recorded.
The results obtained are summarized in Table~\ref{tab_comparison_larger_N}.

From our analysis, it appears that the configurations of minimal
energy reached by the $\cav$ and the $\met$ algorithm are essentially equivalent.

However, on our machine and with our implementation, for the same number of elementary
operations to be performed (spin flips) the PCA algorithm proved to be faster
(about 5 times on the larger instances considered) than the Metropolis.
This is due, essentially, to the parallel nature
of the algorithm that allows (at least on instances of the sizes
taken into account) for a more effective use (by the numerical library)
of the available cores in the computation
of the effect of the spin flips on the energy of the configuration.

Clearly, the time actually spent by each algorithm to find a solution depends
heavily on the implementation and on the system architecture, therefore our claim
on the speed of $\cav$ compared to that of $\met$ should be interpreted cautiously.
However this is an indication that parallel stochastic dynamics may represent a powerful
computational tool to investigate the features of disordered systems in the low temperature regime.
Hence, it seems that a careful implementation on heavily parallel architecture, for instance
GPU, of the PCA proposed in this paper could be really worthwhile. This will be the subject
of future investigations

Even though the size of the sample taken into account 
is rather small,
it is possible to see that the value found by both heuristic algorithms
are very close to what we expect to be the typical value of $m_{N}$.

For what concerns the results obtained with the greedy algorithm $\gre$, they are, as expected,
quite far from those obtained with the two heuristic algorithms. This fact can be interpreted
as a further indication of the complexity of the energy landscape induced by \eqref{eq:hamiltonian}.

\begin{landscape}
\begin{table}[]
\centering
\begin{tabular}{@{}rlllllll@{}}
\toprule
\multicolumn{1}{c}{$N$} & \multicolumn{1}{c}{Instance id} & \multicolumn{1}{c}{$\cav$} & \multicolumn{1}{c}{Avg($\cav$)} & \multicolumn{1}{c}{$\met$} & \multicolumn{1}{c}{Avg($\cav$)} & \multicolumn{1}{c}{$\gre$} & \multicolumn{1}{c}{Avg($\gre$)} \\ \midrule
500                     & 500a                            & 0.415129916               & 0.416601712                    & 0.415129916               & 0.416601712                    & 0.374712619               & 0.387207284                    \\
                        & 500b                            & 0.424031186               &                                 & 0.424031186               &                                 & 0.404622167               &                                 \\
                        & 500c                            & 0.426145394               &                                 & 0.426145394               &                                 & 0.389251214               &                                 \\
                        & 500d                            & 0.401212278               &                                 & 0.401212278               &                                 & 0.38480766                &                                 \\
                        & 500e                            & 0.416489785               &                                 & 0.416489785               &                                 & 0.382642757               &                                 \\ \midrule
1000                    & 1000a                           & 0.414470925               & 0.414146305                    & 0.414470925               & 0.414063509                    & 0.382847286               & 0.384739348                    \\
                        & 1000b                           & 0.412802104               &                                 & 0.412802104               &                                 & 0.38714403                &                                 \\
                        & 1000c                           & 0.426520109               &                                 & 0.426520109               &                                 & 0.383765026               &                                 \\
                        & 1000d                           & 0.40457611                &                                 & 0.404162128               &                                 & 0.371948413               &                                 \\
                        & 1000e                           & 0.412362279               &                                 & 0.412362279               &                                 & 0.397991987               &                                 \\ \midrule
2000                    & 2000a                           & 0.424186053               & 0.420541934                    & 0.424186053               & 0.420515342                    & 0.373116616               & 0.39209371                     \\
                        & 2000b                           & 0.416673303               &                                 & 0.416588939               &                                 & 0.398091317               &                                 \\
                        & 2000c                           & 0.425063334               &                                 & 0.42502059                &                                 & 0.400391099               &                                 \\
                        & 2000d                           & 0.427618814               &                                 & 0.427685529               &                                 & 0.397286458               &                                 \\
                        & 2000e                           & 0.409168165               &                                 & 0.409095598               &                                 & 0.391583062               &                                 \\ \midrule
4000                    & 4000a                           & 0.424745169               & 0.420196116                    & 0.42479645                & 0.420159449                    & 0.401106215               & 0.391286694                    \\
                        & 4000b                           & 0.41805638                &                                 & 0.417861999               &                                 & 0.390199173               &                                 \\
                        & 4000c                           & 0.41629515                &                                 & 0.416234693               &                                 & 0.38404253                &                                 \\
                        & 4000d                           & 0.415214004               &                                 & 0.415233809               &                                 & 0.395550324               &                                 \\
                        & 4000e                           & 0.426669876               &                                 & 0.426670296               &                                 & 0.385535228               &                                 \\ \midrule
8000                    & 8000a                           & 0.416367988               & 0.419943457                    & 0.416174887               & 0.419967205                    & 0.385148017               & 0.391458682                    \\
                        & 8000b                           & 0.414290329               &                                 & 0.414515893               &                                 & 0.395372657               &                                 \\
                        & 8000c                           & 0.424308127               &                                 & 0.424370952               &                                 & 0.392280731               &                                 \\
                        & 8000d                           & 0.421539704               &                                 & 0.421421773               &                                 & 0.397760709               &                                 \\
                        & 8000e                           & 0.423211138               &                                 & 0.42335252                &                                 & 0.386731299               &                                 \\ \bottomrule
\end{tabular}
\caption{Best values of $m_{N}$ found with the PCA algorithms, the Metropolis algorithm and the Greedy
algorithm for instances of various size. Averages of the best values of $m_{N}$ are computed over the instances with the same size.}
\label{tab_comparison_larger_N}
\end{table}
\end{landscape}


%



{\bf Acknowledgements: }
The authors are greateful to Pierre Picco for many fruitful discussions
and to Fabio Lucio Toninelli for his kind help.
AT thanks the Department of Mathematics of the University of Rome
``Tor Vergata'' for the support received and for the warm hospitality.
Our work has been partially supported by PRIN 2012, Problemi matematici in teoria cinetica ed
applicazioni.
BS thanks
the support of the A*MIDEX project (n. ANR-11-IDEX-0001-02) funded by the  ``Investissements d'Avenir" French Government program, managed by the French National Research Agency (ANR).
The authors want to thank the anonymous referees for their precious work.

\end{document}